\documentclass[sigconf,10pt,nonacm]{acmart}

\usepackage{xurl}
\usepackage{hyperref}

\usepackage{booktabs}

\usepackage{tikz}
\usetikzlibrary{positioning,matrix,fit,shapes,backgrounds,patterns,arrows, arrows.meta, calc, decorations.pathreplacing}
\usetikzlibrary{arrows,shapes,decorations,automata,backgrounds,petri,bending}

\usepackage{amsmath}
\usepackage{amsthm}
\usepackage{amsfonts}

\theoremstyle{definition}
\newtheorem{theorem}{Theorem}[section] 
\newtheorem{definition}{Definition}[section]
\newtheorem{corollary}{Corollary}[theorem]

\usepackage{cleveref}  

\usepackage{mathtools, nccmath}

\usepackage{courier}
\usepackage{multirow}
\usepackage{subcaption} 
\usepackage{algorithm}
\usepackage[noend]{algpseudocode}
\usepackage{xspace}
\usepackage{svg}

\usepackage{bm} 

\usepackage[group-separator={,},group-minimum-digits=4]{siunitx} 

\usepackage{listings}

\usepackage{hhline}

\usepackage{pgfplots}

\definecolor{dkgreen}{rgb}{0,0.6,0}
\definecolor{gray}{rgb}{0.5,0.5,0.5}
\definecolor{mauve}{rgb}{0.58,0,0.82}
\definecolor{lightgray}{gray}{0.90}
\definecolor{ms-office-blue}{rgb}{0.36, 0.60, 0.83 } 

\usepackage{array}
\newcolumntype{P}[1]{>{\centering\arraybackslash}p{#1}}  
\newcolumntype{M}[1]{>{\centering\arraybackslash}m{#1}} 

\usepackage[nomessages]{fp}

\newcommand{\verisplit}{VeriSplit\xspace}
\newcommand{\codebaseLinesOfCode}{8728\xspace}

\newcommand*{\circled}[1]{\tikz[baseline=(char.base)]{
            \node[shape=circle,draw,inner sep=1pt,scale=0.8] (char) {#1};}\xspace} 

\newcommand*\emptycirc[1][1ex]{\tikz\draw (0,0) circle (#1);} 
\newcommand*\halfcirc[1][1ex]{%
  \begin{tikzpicture}
  \draw[fill] (0,0)-- (90:#1) arc (90:270:#1) -- cycle ;
  \draw (0,0) circle (#1);
  \end{tikzpicture}}
\newcommand*\fullcirc[1][1ex]{\tikz\fill (0,0) circle (#1);}

\newcommand{\iters}{50}

\newcommand{\vitLatencyReduceUB}{$83\%$\xspace}
\newcommand{\vitLatencyReduceLB}{$28\%$\xspace}

\begin{document}

\date{}

\title{\verisplit{}: Secure and Practical Offloading of Machine Learning Inferences across IoT Devices}


\author{Han Zhang}
\affiliation{
  \institution{Carnegie Mellon University}
  \city{Pittsburgh}
  \state{Pennsylvania}
  \country{USA}
}
\author{Zifan Wang}
\affiliation{
  \institution{Carnegie Mellon University}
  \city{Pittsburgh}
  \state{Pennsylvania}
  \country{USA}
}
\author{Mihir Dhamankar}
\affiliation{
  \institution{Carnegie Mellon University}
  \city{Pittsburgh}
  \state{Pennsylvania}
  \country{USA}
}
\author{Matt Fredrikson}
\affiliation{
  \institution{Carnegie Mellon University}
  \city{Pittsburgh}
  \state{Pennsylvania}
  \country{USA}
}
\author{Yuvraj Agarwal}
\affiliation{
  \institution{Carnegie Mellon University}
  \city{Pittsburgh}
  \state{Pennsylvania}
  \country{USA}
}

\begin{abstract}
Many Internet-of-Things (IoT) devices rely on cloud computation resources to perform machine learning inferences. This is expensive and may raise privacy concerns for users. Consumers of these devices often have hardware such as gaming consoles and PCs with graphics accelerators that are capable of performing these computations, which may be left idle for significant periods of time.
While this presents a compelling potential alternative to cloud offloading, concerns about the integrity of inferences, the confidentiality of model parameters, and the privacy of users' data mean that device vendors may be hesitant to offload their inferences to a platform managed by another manufacturer.

We propose \verisplit{}\footnote{\url{https://github.com/synergylabs/VeriSplit}}, a framework for offloading machine learning inferences to locally-available devices that address these concerns. We introduce masking techniques to protect data privacy and model confidentiality, and a commitment-based verification protocol to address integrity. Unlike much prior work aimed at addressing these issues, our approach does not rely on computation over finite field elements, which may interfere with floating-point computation supports on hardware accelerators and require modification to existing models.
We implemented a prototype of \verisplit{} and our evaluation results show that, compared to performing computation locally, our secure and private offloading solution can reduce inference latency by \vitLatencyReduceLB{}--\vitLatencyReduceUB{}. 

\end{abstract}
\settopmatter{printfolios=true}

\maketitle

\section{Introduction}

Many Internet-of-Things (IoT) devices now support diverse functionality enabling the vision of smart and connected homes (e.g., security cameras~\cite{website-wyze-cam} and video doorbells~\cite{website-blink-doorbell}). Often these devices come with limited hardware and are inexpensive to purchase. Given their limited computing capabilities, device vendors set up dedicated cloud services to support advanced services like machine learning inferences as additional subscriptions~\cite{website-wyze-case-study, website-wyze-cam-subscription}. Since cloud services can incur high operational costs, researchers have proposed techniques to minimize works to be offloaded, such as executing part of the model on the local devices~\cite{zhao2018deepthings, kang2017neurosurgeon}.

Meanwhile, smart home users might have other devices in their homes, such as gaming consoles, computers and mobile devices, and high-end IoT devices with expensive hardware (e.g., vacuum robots with GPUs~\cite{news-roomba-qualcomm-chips}). These devices are wall-powered, have significant computation resources, and are often idle since they are used sporadically. 
When these devices are less busy, why not have resource-limited IoT devices offload certain computation tasks to them? Lower-end IoT devices could enlist the help of powerful local machines and potentially reduce their cloud service costs. Ultimately, these savings can be passed down to smart home users, enabling IoT vendors to charge aggressively yet remain profitable. 
Prior works have studied the challenges of general-purpose computation offload in mobile devices (e.g., \cite{cuervo2010maui, chun2011clonecloud}). In this paper, we want to focus on offloading machine learning inference applications not only because of their increasing popularity in IoT settings \cite{boovaraghavan2021mliot,synthetic-sensors,laput2018ubicoustics, merenda2020edge} but also to leverage unique characteristics of ML workloads to develop specialized solutions for strong security and practicality.

Specifically, there are three major security and practicality concerns we must address to convince IoT device vendors to offload computation to other devices, especially when those devices are controlled by third parties.
First, it is necessary to provide input data with third-party workers for computation. Offloading devices must preserve the \textbf{data privacy} if users do not want to share their data with third parties. 
Second, offloading devices must also provide the model weights to workers, which may be proprietary and raise concerns of \textbf{model confidentiality}. 
Finally, ensuring the \textbf{integrity} of the offloaded inferences is critical~\cite{papernot2016sok-integrity-definition}. Otherwise, cheating workers can perform inferences with lightweight models to save computation costs and remain undetected.

Many recent research works have been proposed for secure offloading of machine learning inferences, such as designing cryptographic protocols for data privacy~\cite{gilad2016cryptonets, juvekar2018gazelle, huang2022cheetah, mishra2020delphi}, applying homomorphic encryption for model confidentiality~\cite{jiang2018edm}, and leveraging verifiable computation to provide inference integrity~\cite{tramer2018slalom, ghodsi2017safetynets, liu2021zkcnn, weng2021mystique}.
These approaches, while indeed promising in providing strong security guarantees, often involve heavyweight computation and high communication overhead, diminishing the values of computation offloading and the practicality of such solutions (e.g., secure inference on ResNet-32 raises the latency from hundreds of milliseconds to 3.8 seconds~\cite{mishra2020delphi}, and some assume high-speed communication channel like PCI-e bus~\cite{tramer2018slalom}). Moreover, cryptographic approaches usually require finite field arithmetic, affecting the model's quality (such as reducing accuracy when models are initially trained with floating point arithmetic operations)~\cite{tramer2018slalom}.

In this paper, we present \verisplit{}, a new offloading framework for IoT devices that addresses the limitations of prior approaches, namely practicality and incompatibility with floating point models, while providing strong security and privacy guarantees for offloading with untrusted third-party workers.
First, \verisplit{} protects data privacy so that the offloading device does not have to share users' private data with workers by applying masks to input data and removing masks with pre-computed values.
Second, \verisplit{} preserves model confidentiality, so the offloading device doesn't have to reveal model parameters by masking parameters and leveraging multiple works to cancel masked weights. 
Third, \verisplit{} ensures inference integrity using a commit-based verifiable offloading algorithm, which supports partial and asynchronous verification to reduce the performance overhead by removing it from the critical path for inferences. 
Finally, unlike prior work that requires finite field arithmetic for perfect security, \verisplit{} can be compatible with floating point arithmetic with the relaxation of the security properties. Keeping models in floating points gives device vendors more flexibility in fine-tuning their models to achieve better performances (e.g., higher accuracy) compared to finite field conversions~\cite{tramer2018slalom}. Still, our security analysis provides concrete proofs and empirical results of \verisplit{}'s strong security guarantees.

We implement a prototype of \verisplit{} and evaluate it with different ML models, as well as with various security and privacy options enabled. Our evaluation results show that, compared to local execution on the IoT device itself, \verisplit{} can reduce inference latency by \vitLatencyReduceLB{}--\vitLatencyReduceUB{}, depending on the desired security guarantees and device configurations.
Therefore, \verisplit{}'s low overhead makes it a practical and secure solution for IoT devices with limited resources to execute large ML models without needing first-party cloud services.

In summary, we make the following contributions:
\begin{itemize}
    \item We identify an emerging use case for offloading machine learning inferences from IoT devices to other local machines from different vendors. We identify critical challenges in realizing this vision and the limitations of existing and simple baseline approaches.
    \item We propose \verisplit{}, a comprehensive framework to facilitate practical cross-device offloading while providing solutions for computation integrity, data privacy, and model confidentiality. \verisplit{} also addresses new challenges to remain compatible with floating point operations instead of requiring finite field arithmetic. 
    \item We implement a prototype of \verisplit{} and evaluate various offloading scenarios. Our results show that \verisplit{} offloading can reduce inference latency by \vitLatencyReduceLB{}--\vitLatencyReduceLB{} while providing strong security and privacy guarantees. 
    
\end{itemize}

\section{Related Work}
\label{sec:related-works}

\begin{table*}[t]
    \centering
    \small
    
\begin{tabular}{M{60mm}M{15mm}M{20mm}M{15mm}M{20mm}M{20mm}}
   \textbf{Approaches}  & \textbf{Data Privacy} & \textbf{Model Confidentiality}  & \textbf{Integrity} & \textbf{Floating Point Models} & \textbf{No Special Hardware on Workers} \\
   \midrule
   Efficient Verification Algorithms (e.g., SafetyNets\cite{ghodsi2017safetynets}, VeriML~\cite{zhao2021veriml}) & \emptycirc & \emptycirc & \fullcirc & \emptycirc & \fullcirc \\
   \midrule
   Cryptographic Protocols (e.g., Zero-Knowledge Proof~\cite{weng2021mystique}, Homomorphic Encryption~\cite{jiang2018edm})  & \fullcirc & \fullcirc & \emptycirc & \emptycirc & \fullcirc \\
   \midrule
    Trusted Execution Environments (e.g., Slalom~\cite{tramer2018slalom}, Graviton~\cite{volos2018graviton})  & \fullcirc & \halfcirc &  \fullcirc & \halfcirc & \emptycirc \\
   \midrule
   \textbf{\verisplit{} (Ours)} & \fullcirc & \fullcirc & \fullcirc & \fullcirc & \fullcirc \\
   \bottomrule
\end{tabular}

    \caption{Comparison of \verisplit{} and other related works. Systems relying on Trusted Execution Environments (TEEs) require workers to have special hardware. Depending on the configuration (CPUs vs. GPUs), it may be possible to protect model confidentiality while using floating-point models. Meanwhile, verifiable and cryptographic models require finite field operations. In comparison, \verisplit{} provides all three goals without special hardware requirements and keeps models in floating points.}
    \label{tab:related-work-comparison}
\end{table*}

Several research efforts have focused on addressing different challenges with secure offloading of ML inferences. \Cref{tab:related-work-comparison} compares \verisplit{} with the most relevant approaches, as we discuss in the section. At a high level, a major advantage of \verisplit{} is its compatibility with floating point operations, unlike prior works that rely on finite field arithmetic for perfect security but sacrifice model quality (e.g., accuracy) and inference speed. Supporting floating point operations introduces new challenges like numerical errors and precision issues. We provide quantitative analyses with various floating point settings to help vendors decide security-accuracy tradeoffs during deployment.

\paragraph{Trusted Hardware}
To facilitate secure and verifiable computation offload, several prior works leverage hardware support using trusted execution environments (TEEs). 
Slalom \cite{tramer2018slalom} utilizes a CPU TEE on the worker device (e.g., Intel secure enclaves) to ensure the correctness of ML inferences on GPUs. Similarly, DeepAttest~\cite{chen2019deepattest} relies on TEEs to attest the integrity of DNN models. TrustFL~\cite{zhang2020TrustFL} and PPFL~\cite{mo2021ppfl} also leverage TEEs (on untrusted devices) to ensure the integrity of federated learning based training. 

Several recent efforts propose new secure accelerators ~\cite{volos2018graviton,zhu2020rack-scale-tee,hua2022guardnn} to expand the trusted computing base beyond the CPU. Notably, Graviton~\cite{volos2018graviton} brings trusted execution environments to GPUs, enabling new applications such as privacy-preserving, high-performance video analysis~\cite{poddar2020visor}. Efforts like SecureTF~\cite{quoc2020securetf} extend ML frameworks with TEE support to help with developer adoption. Commercial efforts~\cite{website-nvidia-confidential} have started adding support for TEEs on datacenter GPUs. 

With a trusted computing base on workers, it should be possible to achieve all three goals by running model inferences inside TEEs. However, several practical challenges hinder its adoption for IoT devices. First, executing inferences with TEEs inside CPUs incurs high performance overhead~\cite{tramer2018slalom}. Furthermore, TEE-enabled accelerators are too costly to be widely deployed in users' smart homes (e.g., datacenter GPUs cost over \$10,000~\cite{news-nvidia-hopper-price}).  Instead, \verisplit{} provides a secure and practical solution without requiring any trusted hardware, enabling us to choose a variety of IoT devices for offloading. 

\paragraph{Efficient Verification}
Various techniques for general-purpose verifiable computation~\cite{gennaro2010VC} and proof techniques for practical verification\cite{costello2015geppetto, parno2013pinocchio} have been used to support verifiable inferences using ML models.
For example, SafetyNets~\cite{ghodsi2017safetynets} proposes interactive proof (IP) protocols and VeriML~\cite{zhao2021veriml} leverages succinct non-interactive arguments of knowledge (SNARK) proofs to make the inference execution verifiable. Unfortunately, these proof generation techniques require computing over finite fields (instead of real numbers) and do not support non-linear activation functions (and hence must use quadratic activation approximation). Therefore, these approaches do not scale to large neural networks while preserving high accuracy of the original, unmodified models. \verisplit{} avoids this limitation by keeping the original neural network models without modification.

In addition to generating succinct proofs, prior works have explored other techniques for efficient verification. Fiore et al.~\cite{fiore2016hash-argue} propose the approach of ``hash-first, verify-later'' to verify computations on outsourced data, similar to \verisplit{}'s asynchronous verification design. Zhang and Muhr~\cite{zhang2021tee-selective-testing} introduce a selective testing mechanism for federated learning that also use Merkle tree hashes to record intermediate results.  GOAT~\cite{asvadishirehjini2020goat} performs probabilistic verification of the intermediate results of machine learning inferences. While \verisplit{} builds upon the idea of probabilistic verification, we enable partial verification of the intermediate results with a novel Merkle tree hash design. 

\paragraph{Cryptography for Machine Learning Applications}
Many recent works propose novel cryptographic protocols for secure ML inferences. These solutions ensure the integrity of the computation since unfaithful executions can be easily detected by the client (e.g., resulting in corrupted ciphertext). One popular approach is by using partial or fully homomorphic encryption (FHE) for matrix multiplication used for inferences~\cite{xu2020secureDL, gilad2016cryptonets, aono2017additive-homomorphic-encryption, jiang2018edm}. In addition to providing inference integrity, homomorphic encryption often preserves data privacy from untrusted workers because it enables computation over encrypted data. Despite recent advances in making FHE more efficient~\cite{riazi2020heax} and using accelerators~\cite{samardzic2021f1}, homomorphic encryption still imposes a significant performance penalty, slowing down inferences by orders of magnitudes. 

To alleviate FHE's overhead, recent works have explored incorporating secure multi-party computation into the offloading protocol design~\cite{mohassel2017secureml} or with homomorphic encryption~\cite{juvekar2018gazelle, mishra2020delphi}. In addition, systems like MiniONN~\cite{liu2017minionn} leverage oblivious transfers to prevent private data from leaving the device while keeping model secrets on the worker machine. 

Compared to \verisplit{}, these techniques incur high computation overhead, especially challenging for low-cost IoT devices. Moreover, these techniques still require computation over finite fields and are incompatible with non-linear activation functions of a neural network. Approximating activation with quadratic functions often impacts model privacy and sometimes affects model training convergence~\cite{mishra2020delphi}. \verisplit{}, in contrast, supports offloading unmodified, state-of-the-art, neural network models on IoT devices. 

\section{\verisplit{} Overview}
\label{sec:overview}

\begin{figure}
    \centering
    \input{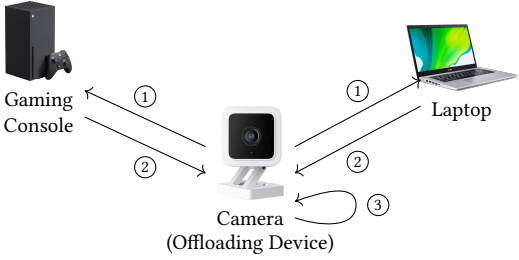}
    \caption{\verisplit{} deployment example in a smart home. The IoT device (a smart camera) can offload ML inferences to one or more local devices by sharing input data and model parameters (\protect\circled{1}) and receiving results (\protect\circled{2}). Later, the camera can verify the integrity of the inference by repeating selected computations and comparing the results (\protect\circled{3}), doing it asynchronously to avoid additional latency during inference.}
    \label{fig:overview}
\end{figure}

In this section, we start with a high-level overview of \verisplit{}, followed by our detailed design goals (\Cref{sec:design-goals}) and our threat model (\Cref{sec:threat-model}). The next few sections address solutions to each design goal in detail (\Cref{sec:data-privacy,sec:model-confidentiality,sec:inference-integrity}).

\Cref{fig:overview} illustrates the overall workflow of \verisplit{} offloading across local devices using a resource-constrained IoT device (a smart camera in this example). As the offloading device, it first goes through setup processes with local devices (not shown in the figure). This setup process includes various steps depending on the requirements (data privacy, confidentiality, and integrity), and the offloading device shares model weights with workers.

During inference, the camera shares input data with one or more workers (Step \protect\circled{1}). \verisplit{} needs multiple workers to support model confidentiality. To support only data privacy or integrity, a single worker is sufficient. Then, workers respond with the inference results and additional proofs for integrity, if applicable (Step \protect\circled{2}). Next, to check the integrity of the offloaded computation, the camera can decide when and how many results to verify thanks to \verisplit{}'s asynchronous and partial verification design.

We integrate two types of offloading solutions to accommodate various model architectures and desired security and privacy properties in \verisplit{}:
\begin{itemize}
    \item \textbf{Holistic Offload.} The most straightforward approach is to offload the model in its entirety. The device shares all model parameters with workers and sends input data during inference. After executing the full model, workers return the final predictions. Holistic offloading greatly improves performance but can only ensure inference integrity.
    
    \item \textbf{Layer-by-Layer Offload.} To protect data privacy and model confidentiality, we propose a layer-by-layer offloading alternative due to non-linearity across layers. The device can choose which layers to offload independently. Due to its added communication overhead, this approach may not be ideal for certain use cases or network conditions. 
\end{itemize}

\subsection{Design Goals}
\label{sec:design-goals}

We want to propose a practical solution to motivate IoT device vendors to enlist help from other co-located devices in the same user's home. Therefore, we must address many common concerns related to data privacy, model confidentiality, computation integrity, and performance overhead from the \textbf{offloading device vendor's} perspective. 

\paragraph{Protecting User Data from Third-Party Workers}
It is necessary for the offloading device to share input data with workers. However, this process inevitably leaks smart home users' private data to third parties. If the user does not want to expose their information to other local devices, \verisplit{} must support data privacy while preserving the functionality and accuracy of the offloaded computation. 

\paragraph{Preserving Model Secrets}
Device vendors may hesitate, or even refuse, to share their models with third-party workers due to intellectual property concerns. To protect vendors' models, \verisplit{} must provide practical solutions that prevent third-party workers' unauthorized use of secret models. 

\paragraph{Ensuring Correct Computation}
A fundamental requirement for offloading is to ensure the correctness of the computation. Otherwise, lazy workers may avoid work and return arbitrary results. Therefore, \verisplit{} should provide efficient verification solutions  so that device vendors can easily check the results reported by workers are computed correctly. 

\paragraph{Practical and Low Overhead}
While addressing the aforementioned secure and private offloading features, \verisplit{} needs to have low overhead to be practical for different IoT device use cases. We provide comprehensive analyses regarding \verisplit{}'s performance trade-offs.

\subsection{Threat Model}
\label{sec:threat-model}

We assume workers performing the computation are untrusted. They may attempt to steal any data the offloading device shares, such as the inputs for inference and the ML model parameters (e.g., layer weights and bias), raising concerns for data privacy and model confidentiality. In addition, workers may have incentives to ``cheat'' by executing inferences using lightweight models or reporting incorrect results to save computation costs or reduce application accuracy from competing IoT vendors. Therefore, it calls for verification solutions to ensure inference integrity. We assume multiple workers used in the \emph{same} round of inference offloading are non-colluding (e.g., belonging to different vendors) as a prerequisite for model confidentiality. 
Non-colluding workers are common assumptions in many prior works \cite{mohassel2017secureml, riazi2018chameleon}.

Such a strong attacker model may be too restrictive in certain cases and can be relaxed to reduce the performance overhead, depending on the use cases and deployment environment. Specifically, if the smart home user can trust the worker devices with their personal data (e.g., if they use their own PC as workers), they may not require data privacy protection. However, model confidentiality and integrity are still important for \emph{offloading device} vendors. On the other hand, if the offloading device only cares about integrity (e.g., when using public models for feature extraction and keeping proprietary layers on-device, as used in transfer learning applications), they can further relax the requirement of model confidentiality, further reducing \verisplit{}'s overhead.

\section{Data Privacy}
\label{sec:data-privacy}

\begin{figure}[t]
    \centering
    \begin{tikzpicture}[every node/.style={font=\footnotesize, inner sep=2pt, fill=none, 
    text=black, align=center}]
	
	\tikzstyle{dualline} = [draw, {latex}-{latex}]
	\tikzstyle{line} = [draw, -{latex}]
		
	\def\spacing{18pt}
	
	\def\deviceX{0pt}
	\def\deviceY{0pt}
	\def\workerX{140pt}
	\def\workerY{0pt}
	
	\def\vLineStart{-5pt}
	\def\vLineEnd{-100pt}
	
	\node[draw,fill=white, rounded corners] (device) at (\deviceX, \deviceY) {Offloading Device};
	\node[draw,fill=white, rounded corners] (worker) at (\workerX, \workerY) {Worker};

	\draw[color=ms-office-blue, very thick] (\deviceX, \vLineStart) -- (\deviceX, \vLineEnd);
	\draw[color=ms-office-blue, very thick] (\workerX, \vLineStart) -- (\workerX, \vLineEnd);

        \def\modelY{\vLineStart-0.5*\spacing};
	\path[line] (\deviceX, \modelY) --
	node[above]{Model $W, b$}
	(\workerX, \modelY);
 
	\def\genNoiseY{\modelY- 0.5*\spacing};
	\node[draw,fill=white] (gen-noise) at (\deviceX, \genNoiseY) {Generate random mask: $\epsilon_i$};
	
        \def\calcWe{\genNoiseY- 0.75*\spacing};
	\node[draw,fill=white] (calc-We) at (\deviceX, \calcWe) {Precompute: $W\epsilon_i$};

        \def\sepLineY{\calcWe-0.5*\spacing}
        \draw[dashed, orange, very thick] (\deviceX-50, \sepLineY) -- 
        (\workerX+50, \sepLineY)
        node[above left]{\textbf{Setup}}
        node[below left]{\textbf{Inference}};

        \def\reqY{\sepLineY-0.75*\spacing};
	\path[line] (\deviceX, \reqY) --
	node[above]{\quad Data ($x_i+\epsilon_i$)}
	(\workerX, \reqY);

        \def\calcWxeb{\reqY- 0.75*\spacing};
	\node[draw,fill=white] (calc-Wxeb) at (\workerX, \calcWxeb) {$y' = W(x_i+\epsilon_i)+b$};

        \def\respY{\calcWxeb-0.7*\spacing}; 
	\path[line] (\workerX, \respY) --
	node [above] {Result $y'$}
	(\deviceX, \respY);

        \def\calcWxb{\respY- 0.7*\spacing};
	\node[draw,fill=white] (calc-Wxb) at (\deviceX, \calcWxb) {$y = y' - W\epsilon_i = W x_i + b$};

\end{tikzpicture}
    \caption{\verisplit{}'s workflow for data privacy. During setup, the device shares model parameters with the worker, generates multiple one-time masks $\epsilon_i$, and precomputes their values $W \epsilon_i$. For each inference, the device sends masked input $x+\epsilon_i$ and then subtracts the precomputed values from the received result $y'$ to recover actual computation results.}
    \label{fig:workflow-data-privacy}
\end{figure}
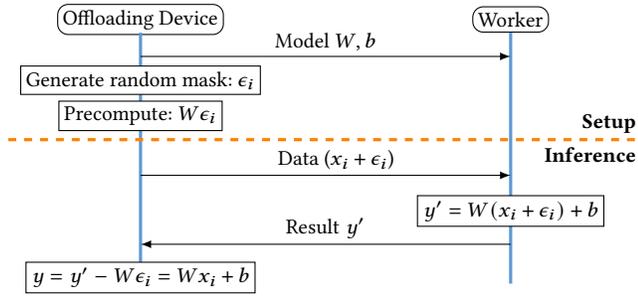

This section discusses how \verisplit{} utilizes linear arithmetic and one-time masks to preserve users' private data during offloading. We do this to prevent third-party workers from accessing user data while providing inferences. 

Prior research has proposed using cryptography for data privacy in inference offloading (e.g., garbled circuits, homomorphic encryption, etc.). These approaches are still heavyweight, despite recent advances significantly improving performance~\cite{juvekar2018gazelle, mishra2020delphi}. For example, Delphi~\cite{mishra2020delphi} takes 3.8 seconds for a single secure inference of ResNet-32, while similar models only need a few hundred milliseconds\footnote{Baselines empirically measured by us since not presented in Delphi~\cite{mishra2020delphi}.}. 

We introduce a linear masking solution to protect data privacy in \verisplit{} without relying on computationally expensive cryptographic primitives. It must be used with layer-by-layer offloading due to non-linearity across layers (e.g., activation functions). Our masking design is partly inspired by several prior cryptographic inference protocols \cite{tramer2018slalom, mishra2020delphi}; we provide additional security analysis for floating-point masks (\Cref{sec:security-analysis}) instead of finite field.

\Cref{fig:workflow-data-privacy} presents the workflow of \verisplit{}'s data privacy protection. During setup, the offloading device shares model parameters (e.g., weights $W$ and bias $b$) with the worker and locally generates one-time masks ($\epsilon_i$). 
In addition, the device computes the value $W \epsilon_i$ for each mask and stores it locally. 
It can generate many masks (and precompute $W \epsilon_i$'s)  during idle times since each inference will consume a fresh one. This pre-processing step is commonly used by prior works~\cite{tramer2018slalom, mishra2020delphi}.

For a particular inference $i$, the device sends masked inputs ($x' = x + \epsilon_i$) instead of the real data. The worker executes inference and replies with the results ($y' = W (x + \epsilon_i) + b$). The worker never has access to users' private data. Once the device receives $y'$, it can subtract the precomputed mask noise ($W \epsilon_i$) and obtain the true layer output $y = y' - W\epsilon_i = Wx_i+b$. 
This solution offloads the expensive matrix multiplication computation to workers \emph{during inferences}. As a compromise, the device must perform pre-computation ($W \epsilon_i$) asynchronously during setup or obtain pre-computed mask values from a trusted source (e.g., vendor backend can generate many masks and distribute them to edge devices).

\section{Model Confidentiality}
\label{sec:model-confidentiality}

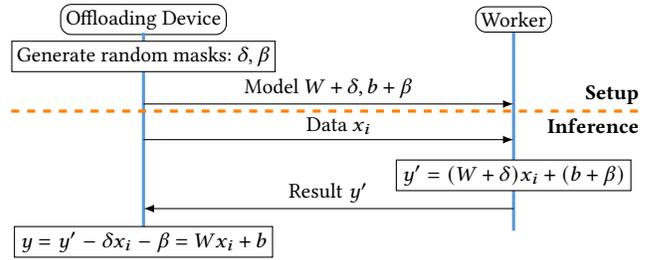
\begin{figure}[t]
    \centering
    \begin{tikzpicture}[every node/.style={font=\footnotesize, inner sep=2pt, fill=none, 
    text=black, align=center}]
	
	\tikzstyle{dualline} = [draw, {latex}-{latex}]
	\tikzstyle{line} = [draw, -{latex}]
		
	\def\spacing{18pt}
	
	\def\deviceX{0pt}
	\def\deviceY{0pt}
	\def\workerX{140pt}
	\def\workerY{0pt}
	
	\def\vLineStart{-5pt}
	\def\vLineEnd{-80pt}
	
	\node[draw,fill=white, rounded corners] (device) at (\deviceX, \deviceY) {Offloading Device};
	\node[draw,fill=white, rounded corners] (worker) at (\workerX, \workerY) {Worker};

	\draw[color=ms-office-blue, very thick] (\deviceX, \vLineStart) -- (\deviceX, \vLineEnd);
	\draw[color=ms-office-blue, very thick] (\workerX, \vLineStart) -- (\workerX, \vLineEnd);

	\def\genNoiseY{\vLineStart- 0.5*\spacing};
	\node[draw,fill=white] (gen-noise) at (\deviceX, \genNoiseY) {Generate random masks: $\delta, \beta$};

        \def\modelY{\genNoiseY-1*\spacing};
	\path[line] (\deviceX, \modelY) --
	node[above]{Model $W+\delta, b+\beta$}
	(\workerX, \modelY);

        \def\sepLineY{\modelY-0.15*\spacing}
        \draw[dashed, orange, very thick] (\deviceX-50, \sepLineY) -- 
        (\workerX+50, \sepLineY)
        node[above left]{\textbf{Setup}}
        node[below left]{\textbf{Inference}};

        \def\reqY{\sepLineY-0.6*\spacing};
	\path[line] (\deviceX, \reqY) --
	node[above]{\quad Data $x_i$}
	(\workerX, \reqY);

        \def\calcWxeb{\reqY- 0.75*\spacing};
	\node[draw,fill=white] (calc-Wxeb) at (\workerX, \calcWxeb) {$y' = (W+\delta)x_i+(b+\beta)$};

        \def\respY{\calcWxeb-0.7*\spacing}; 
	\path[line] (\workerX, \respY) --
	node [above] {Result $y'$}
	(\deviceX, \respY);

        \def\calcWxb{\respY- 0.7*\spacing};
	\node[draw,fill=white] (calc-Wxb) at (\deviceX, \calcWxb) {$y = y' - \delta x_i - \beta = W x_i + b$};

\end{tikzpicture}
    \caption{A simple but ineffective approach to adding masks for model confidentiality. This does not save any work for the device during inference, since it still needs to compute $\delta x_i$.}
    \label{fig:workflow-incorrect-confidentiality}
\end{figure}

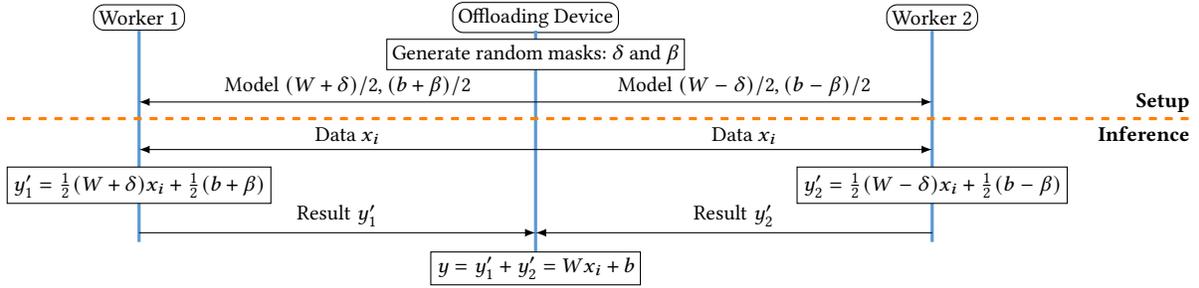
\begin{figure*}[t]
    \centering
    \begin{tikzpicture}[every node/.style={font=\footnotesize, inner sep=2pt, fill=none, 
    text=black, align=center}]
	
	\tikzstyle{dualline} = [draw, {latex}-{latex}]
	\tikzstyle{line} = [draw, -{latex}]
		
	\def\spacing{18pt}
	
	\def\deviceX{0pt}
	\def\deviceY{0pt}
	\def\workerAX{150pt}
	\def\workerAY{0pt}
        \def\workerBX{-150pt}
	\def\workerBY{0pt}
	
	\def\vLineStart{-5pt}
	\def\vLineEnd{-85pt}
	
	\node[draw,fill=white, rounded corners] (device) at (\deviceX, \deviceY) {Offloading Device};
	\node[draw,fill=white, rounded corners] (workerA) at (\workerAX, \workerAY) {Worker 2};
	\node[draw,fill=white, rounded corners] (workerB) at (\workerBX, \workerBY) {Worker 1};

	\draw[color=ms-office-blue, very thick] (\deviceX, \vLineStart) -- (\deviceX, \vLineEnd-10pt);
	\draw[color=ms-office-blue, very thick] (\workerAX, \vLineStart) -- (\workerAX, \vLineEnd);
        \draw[color=ms-office-blue, very thick] (\workerBX, \vLineStart) -- (\workerBX, \vLineEnd);

	\def\genNoiseY{\vLineStart- 0.5*\spacing};
	\node[draw,fill=white] (gen-noise) at (\deviceX, \genNoiseY) {Generate random masks: $\delta$ and $\beta$};

        \def\modelY{\genNoiseY-1*\spacing};
	\path[line] (\deviceX, \modelY) --
	node[above]{\quad Model $(W-\delta)/2, (b-\beta)/2$}
	(\workerAX, \modelY);
        \path[line] (\deviceX, \modelY) --
	node[above]{\quad Model $(W+\delta)/2, (b+\beta)/2$}
	(\workerBX, \modelY);

        \def\sepLineY{\modelY-0.35*\spacing}
        \draw[dashed, orange, very thick] (\workerBX-50, \sepLineY) -- 
        (\workerAX+100, \sepLineY)
        node[above left]{\textbf{Setup}}
        node[below left]{\textbf{Inference}};

        \def\dataY{\sepLineY-0.65*\spacing};
	\path[line] (\deviceX, \dataY) --
	node[above]{\quad Data $x_i$}
	(\workerAX, \dataY);
        \path[line] (\deviceX, \dataY) -- 
        node[above]{\quad Data $x_i$}
        (\workerBX, \dataY);
        
        \def\calcWxeb{\dataY- 0.75*\spacing};
	\node[draw,fill=white] (calc-Wxeb) at (\workerAX, \calcWxeb) {$y'_2 = \frac{1}{2}(W-\delta)x_i + \frac{1}{2}(b-\beta)$};
	\node[draw,fill=white] (calc-Wxeb) at (\workerBX, \calcWxeb) {$y'_1 = \frac{1}{2}(W+\delta)x_i + \frac{1}{2}(b+\beta)$};

        \def\respY{\calcWxeb-1*\spacing}; 
	\path[line] (\workerAX, \respY) --
	node [above] {Result $y'_2$}
	(\deviceX, \respY);
        \path[line] (\workerBX, \respY) --
	node [above] {Result $y'_1$}
	(\deviceX, \respY);

        \def\calcWxb{\respY- 0.75*\spacing};
    \node[draw,fill=white] (calc-Wxb) at (\deviceX, \calcWxb) {$y = y'_1 + y'_2 = W x_i + b$};

\end{tikzpicture}
    \caption{\verisplit{}'s workflow for model confidentiality. During setup, the offloading device generates masking noises ($\delta, \beta$) and shares modified model parameters with two non-colluding workers. For inferences, the device sends data and receives layer results ($y'_1, y'_2$). It can reconstruct the actual results by combining the two results. }
    \label{fig:workflow-model-confidentiality}
\end{figure*}

For model confidentiality, we propose a similar masking approach to that for data privacy. However, simply adding noise to these parameters does not work. \Cref{fig:workflow-incorrect-confidentiality} illustrates this alternative and why it does not work for offloading since the device must still perform large matrix multiplication ($\delta x_i$ in the last step) for each inference. Since this value is input-dependent, the device can not pre-compute it as in the case of data privacy. Offloading is actually worse in this case. 

In \verisplit{}, we propose a solution leveraging multiple workers to provide model confidentiality. By sharing inference tasks with multiple workers, the device can avoid computing matrix multiplications and leverage the performance benefits of more capable workers. As mentioned in our threat model (\Cref{sec:threat-model}), we assume these workers are non-colluding, so they will not collectively compromise model confidentiality. This is a common threat model in many prior works~\cite{riazi2018chameleon, mohassel2017secureml}. To achieve this, the offloading device can choose two workers from different vendors (or one personal device like a PC owned by the smart home user). 

\Cref{fig:workflow-model-confidentiality} illustrates \verisplit{}'s workflow for model confidentiality. The offloading device generates masks for model parameters (e.g., $\delta$ for weights and $\beta$ for biases) and selects two local workers to share masked values. This is a one-time setup effort; subsequent inferences can use the same weights and biases. During the offload of an inference $i$, the device shares the input data with both workers and receives results $y'_1$ and $y'_2$. By combining these two results, the mask terms cancel out and provide the actual result of $Wx_i + b$. On the other hand, both workers only have access to masked weight parameters. Neither one can reconstruct original values without collusion. We elaborate on the security guarantee this method provides in \Cref{sec:security-analysis}.

The online inference process requires two workers to perform an equal workload with different model parameters. Both workers need to have capable hardware. In comparison, the offloading device now only needs to sum up intermediate results ($y'_1$ and $y'_2$), which is much faster than performing matrix multiplication operations locally.

\section{Inference Integrity}
\label{sec:inference-integrity}

This section explains how \verisplit{} ensures the integrity of the offloaded computation. We propose asynchronous verification to remove verification from the critical path of inferences. We also propose a partial verification mechanism to reduce the communication overhead.

\subsection{Layer-by-Layer Offloading}
If the IoT device requires data privacy or model confidentiality, it has to offload neural network computation layer-by-layer because \verisplit{}'s solutions only apply to linear operations (e.g., unsuitable for activation functions between layers). In the layer-by-layer offload scenario,  verifying the integrity of inferences is fairly straightforward. During the layer-by-layer offloading, the device must transmit all intermediate results to workers. Therefore, the offloading device already has the complete inputs and outputs for each layer. To verify integrity, it can repeat the computation locally and check if the results match. As long as the device saves these results (at the expense of extra storage), it can choose when and how much to verify, retaining flexibility for asynchronous and partial verification. Alternatively, they can use \verisplit{}'s Merkle-tree-based hash commits (explained in the next section) to further reduce local storage overhead, although our prototype currently does not implement this function for layer-by-layer offloading due to ample storage.

\subsection{Holistic Offloading}
If the offloading device only requires integrity guarantees without data confidentiality or data privacy, it can perform holistic offloading --- sending the entire model to the worker and directly sharing the inputs to the model. The worker executes all model layers in one pass and returns the inference outputs. To verify the integrity of any layers, the device needs to access the inputs and outputs of the corresponding layer. Sending all intermediate results during inference can incur high communication overhead. To minimize inference latency, \verisplit{} separates verification from the inference process by proposing a commitment-based, asynchronous solution. The device can request intermediate data from previous inferences whenever it has free cycles. In addition, the device can request partial results from arbitrary layers (instead of full-layer results) if it only wants to check part of the results (to reduce verification overhead). 

\subsubsection{Asynchronous Verification}

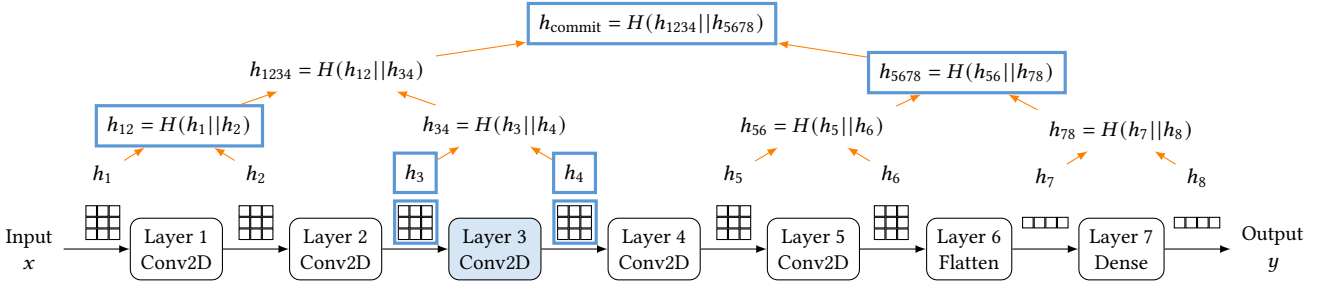
\begin{figure*}[t]
    \centering
    \begin{tikzpicture}[every node/.style={font=\footnotesize, inner sep=4pt, fill=none, 
    text=black, align=center, minimum height=10pt}]

\tikzstyle{dualline} = [draw, {latex}-{latex}]
\tikzstyle{line} = [draw, -{latex}]

\def\layerSpacing{25pt}

\node[anchor=east] at (0, 0) (input) {Input\\$x$};
\node[draw, rounded corners, right=\layerSpacing of input] (layer-1) {Layer $1$\\Conv2D};
\node[draw, rounded corners, right=\layerSpacing of layer-1] (layer-2) {Layer $2$\\Conv2D};
\node[draw, rounded corners, right=\layerSpacing of layer-2, fill=ms-office-blue!25] (layer-3) {Layer $3$\\Conv2D};
\node[draw, rounded corners, right=\layerSpacing of layer-3] (layer-4) {Layer $4$\\Conv2D};
\node[draw, rounded corners, right=\layerSpacing of layer-4] (layer-5) {Layer $5$\\Conv2D};
\node[draw, rounded corners, right=\layerSpacing of layer-5] (layer-6) {Layer $6$\\Flatten};
\node[draw, rounded corners, right=\layerSpacing of layer-6] (layer-7) {Layer $7$\\Dense};
\node[right=\layerSpacing of layer-7] (output) {Output\\$y$};

\path[line] (input.east) --
node[above] {}
(layer-1.west);
\path[line] (layer-1.east) --
node[above] {}
(layer-2.west);
\path[line] (layer-2.east) --
node[above] {}
(layer-3.west);
\path[line] (layer-3.east) --
node[above] {}
(layer-4.west);
\path[line] (layer-4.east) --
node[above] {}
(layer-5.west);
\path[line] (layer-5.east) --
node[above] {}
(layer-6.west);
\path[line] (layer-6.east) --
node[above] {}
(layer-7.west);
\path[line] (layer-7.east) --
node[above] {}
(output.west);

\def\matrixYshift{10pt}
\tikzstyle{matrixCell} = [nodes={fill=white,minimum size=4pt, inner sep=0pt,draw}]

\matrix [matrixCell] 
(m1) at ([yshift=\matrixYshift]$(input)!0.5!(layer-1)$)
{
\node {}; & \node{}; & \node {}; \\
\node {}; & \node{}; & \node {}; \\
\node {}; & \node{}; & \node {}; \\
};
\matrix [matrixCell] 
(m2) at ([yshift=\matrixYshift]$(layer-1)!0.5!(layer-2)$)
{
\node {}; & \node{}; & \node {}; \\
\node {}; & \node{}; & \node {}; \\
\node {}; & \node{}; & \node {}; \\
};
\matrix [matrixCell] 
(m3) at ([yshift=\matrixYshift]$(layer-2)!0.5!(layer-3)$)
{
\node {}; & \node{}; & \node {}; \\
\node {}; & \node{}; & \node {}; \\
\node {}; & \node{}; & \node {}; \\
};
\matrix [matrixCell] 
(m4) at ([yshift=\matrixYshift]$(layer-3)!0.5!(layer-4)$)
{
\node {}; & \node{}; & \node {}; \\
\node {}; & \node{}; & \node {}; \\
\node {}; & \node{}; & \node {}; \\
};
\matrix [matrixCell] 
(m5) at ([yshift=\matrixYshift]$(layer-4)!0.5!(layer-5)$)
{
\node {}; & \node{}; & \node {}; \\
\node {}; & \node{}; & \node {}; \\
\node {}; & \node{}; & \node {}; \\
};
\matrix [matrixCell] 
(m6) at ([yshift=\matrixYshift]$(layer-5)!0.5!(layer-6)$)
{
\node {}; & \node{}; & \node {}; \\
\node {}; & \node{}; & \node {}; \\
\node {}; & \node{}; & \node {}; \\
};
\matrix [matrixCell] 
(m7) at ([yshift=\matrixYshift]$(layer-6)!0.5!(layer-7)$)
{
\node {}; & \node{}; & \node {}; & \node{};\\
};
\matrix [matrixCell] 
(m8) at ([yshift=\matrixYshift]$(layer-7)!0.5!(output)$)
{
\node {}; & \node{}; & \node {}; & \node{};\\
};

\def\hashLevelYshift{10pt}

\node[above=0pt of m1] (h1) {$h_1$};
\node[above=0pt of m2] (h2) {$h_2$};
\node[above=0pt of m3, draw=ms-office-blue, very thick] (h3) {$h_3$};
\node[above=0pt of m4, draw=ms-office-blue, very thick] (h4) {$h_4$};
\node[above=0pt of m5] (h5) {$h_5$};
\node[above=0pt of m6] (h6) {$h_6$};
\node[above=3pt of m7] (h7) {$h_7$};
\node[above=3pt of m8] (h8) {$h_8$};

\node[above=\hashLevelYshift of $(h1)!0.5!(h2)$, draw=ms-office-blue, very thick] (h12) {$h_{12} = H(h_1 || h_2)$};
\node[above=\hashLevelYshift of $(h3)!0.5!(h4)$] (h34) {$h_{34} = H(h_3 || h_4)$};
\node[above=\hashLevelYshift of $(h5)!0.5!(h6)$] (h56) {$h_{56} = H(h_5 || h_6)$};
\node[above=\hashLevelYshift of $(h7)!0.5!(h8)$] (h78) {$h_{78} = H(h_7 || h_8)$};
\path[line, draw=orange] (h1) -- (h12);
\path[line, draw=orange] (h2) -- (h12);
\path[line, draw=orange] (h3) -- (h34);
\path[line, draw=orange] (h4) -- (h34);
\path[line, draw=orange] (h5) -- (h56);
\path[line, draw=orange] (h6) -- (h56);
\path[line, draw=orange] (h7) -- (h78);
\path[line, draw=orange] (h8) -- (h78);

\node[above=13pt of $(h12)!0.5!(h34)$] (h1234) {$h_{1234} = H(h_{12} || h_{34})$};
\node[above=13pt of $(h56)!0.5!(h78)$, draw=ms-office-blue, very thick] (h5678) {$h_{5678} = H(h_{56} || h_{78})$};
\path[line, draw=orange] (h12) -- (h1234);
\path[line, draw=orange] (h34) -- (h1234);
\path[line, draw=orange] (h56.north east) -- (h5678);
\path[line, draw=orange] (h78.north west) -- (h5678);

\node[above=\hashLevelYshift of $(h1234)!0.5!(h5678)$, draw=ms-office-blue, very thick] (hcommit) {$h_{\mathrm{commit}} = H(h_{1234} || h_{5678})$};
\path[line, draw=orange] (h1234) -- (hcommit);
\path[line, draw=orange] (h5678) -- (hcommit);

\draw[draw=ms-office-blue, very thick] 
([xshift=3pt, yshift=3pt]m3.south west) 
rectangle 
++(16pt, 16pt);
\draw[draw=ms-office-blue, very thick] 
([xshift=3pt, yshift=3pt]m4.south west) 
rectangle 
++(16pt, 16pt);

\end{tikzpicture}
    \caption{\verisplit{}'s inference commitment design based on Merkle trees. This example network includes several 2D convolution layers, one flatten and one dense layer. During inference, the worker computes hash values of all layers' intermediate results ($h_1$--$h_8$) and reduces them into the final commit value $h_{\mathrm{commit}}$. Assuming the verifier randomly decides to check results from layer 3 (blue shadow), the worker sends all values in blue boxes as its integrity proof. This commitment mechanism can be expanded to partially verify layer results.}
    \label{fig:construct-merkle-tree}
\end{figure*}

The first key insight of \verisplit{}'s integrity solutions is to move the verification process out of the critical path of the prediction itself while only requiring minimal storage overheads on the worker and verifier (i.e., the offloading device) sides. While prior works can support asynchronous verification (such as incorporating interactive proof techniques into neural network models~\cite{ghodsi2017safetynets}), they require modification to the model and affect its behavior and accuracy. On the other hand, works like Slalom~\cite{tramer2018slalom} require transmitting intermediate results as part of the inference process, incurring high overhead. In addition, the verifier may need to store all intermediate results locally if they want to check them later. In contrast, \verisplit{} verifier only needs to store a single hash commit.

During inferences, the worker saves the input data while discarding all intermediate results, saving on storage. During verification, the verifier asks the worker for the original intermediate values (certain layers' inputs and outputs) and repeats the computation locally to check if the results match. To generate the selected layer results, the worker can repeat the inference a second time (since it stores the input data) and send back the requested values. 

For integrity, \verisplit{} has to check whether the worker provides the same values during verification as those used in the original inferences. For example, a malicious or lazy worker may only execute the correct model if an inference is selected for verification. To address this challenge, we design a Merkle tree-based inference commitment mechanism to capture all intermediate values generated during inference. The verifier receives this short commit value along with the inference results and stores it locally. 
During verification, the verifier can request arbitrary intermediate values from the worker, and the worker must generate proofs showing these values are included in the original commit. 

In \verisplit{}, we designed a Merkle tree construction algorithm to enable flexible verification of arbitrary intermediate values across internal layers of a neural network model.  Merkle tree~\cite{merkle1987tree} data structures have been widely used in many applications for generating short commit messages to provide verifiable integrity guarantees on system states~\cite{andersen2019wave,rogers2007merkle-tree-processor, xing2016intel-sgx-memory, dziembowski2018fairswap}. Recent works have also incorporated them into various machine learning applications~\cite{zhang2021tee-selective-testing, tian2021blockchain-machine-learning}. 

\Cref{fig:construct-merkle-tree} illustrates constructing the Merkle tree commit for an inference. For simplicity, we consider a 7-layer network with five 2D convolution layers, one flatten layer, and one dense layer for final prediction. The sizes of the matrices shown are for illustrative purposes only. During inference, the worker collects all layer inputs and outputs and computes their hash values ($h_1$--$h_8$). Then it generates the inference commit ($h_{\mathrm{commit}}$) by recursively concatenating and hashing values in the Merkle tree. The verifier receives this commit along with the final prediction results at the end of the inference offloading process.

\subsubsection{Partial Verification}

\verisplit{} further extends the Merkle tree construction to support verifying partial layer results at fine granularity. 
For any individual layer results (e.g., a single matrix from \Cref{fig:construct-merkle-tree}), we slice them into multiple non-overlapping, contiguous ``verify units'', covering the entire matrix. We select the number of units based on the desired \emph{verify ratio} --- the minimal amount of results one may want to check at a single round. This way, verifying each unit approximately equals to checking the same ``verify ratio'' of the entire layer results. We empirically select axes for high-dimension matrices depending on the layer types and data shapes. For example, we slice a layer output of shapes $(224, 224, 128)$ with a unit size of $(22, 22, 128)$ for $1\%$ ratio.  

After slicing, we construct a Merkle tree with all units of that layer and use the root hash value as this layer's hash result (i.e., to be used as $h_i$ in \Cref{fig:construct-merkle-tree} for layer $i$). During verification, the verifier requests specific units from a layer and reconstructs the hash commit based on the Merkle tree results to ensure integrity.

\newcommand{\clientSymbol}{$\mathcal{C}$\xspace}
\newcommand{\adversarySymbol}{$\mathcal{A}$\xspace}

\section{Security Analysis}
\label{sec:security-analysis}
In this section, we present a formal security analysis of \verisplit{}'s solutions regarding \verisplit{}'s three main goals (data privacy, model confidentiality, and inference integrity). 

\paragraph{Data Privacy and Model Confidentiality}

As a building block, we define a secure masking scheme between a client \clientSymbol{} and an adversary \adversarySymbol{} that produces indistinguishable results for \adversarySymbol{} when \clientSymbol{} randomly picks a value from choices $(x_1, x_2)$ and applies a mask from range $[-\epsilon, \epsilon]$. We assume both choices for $x$ and mask ranges are known to \adversarySymbol{} and \clientSymbol{}:
\begin{definition}[$\epsilon$-Masking Schemes]\label{def:masking-scheme}
    An $\epsilon$-masking scheme consists of the following steps:
    \begin{itemize}
        \item $x_i \gets (x_1, x_2)$, $\epsilon_i \gets [-\epsilon, \epsilon]$: \clientSymbol{} uniformly randomly selects a value $x_i$ and a mask $\epsilon_i$. \clientSymbol{} shares the masked value $(x_i + \epsilon_i)$ with \adversarySymbol{}.
        \item \adversarySymbol{} guesses whether the selected value is $x_1$ or $x_2$ based on the observation of $(x_i + \epsilon_i)$.
    \end{itemize}
\end{definition}

\Cref{def:masking-scheme} defines an $\epsilon$-bound masking scheme. Ideally, with perfect security, \adversarySymbol{} should not gain any advantage in guessing $x_i$ after observing the masked value of $(x_i + \epsilon_i)$. Therefore, we present the following definition of security:
\begin{definition}[Secure Masking Schemes]\label{def:secure-masking}
    An $\epsilon$-masking scheme is secure if it produces \emph{indistinguishable} results such that \adversarySymbol{} can only succeed at deciding the original value of $x_i$ with up to $50\%$ probability.
\end{definition}

We can further define failures of such schemes as follows:
\begin{definition}[Masking Scheme Failures]\label{def:masking-failures}
    A masking scheme fails when \adversarySymbol{} can distinguish the original value of $x_i$ based on observed $(x_i + \epsilon_i)$ with $>50\%$ success rates.
\end{definition}

Next, following these definitions, we can show that an $\epsilon$-masking scheme can achieve perfect security by selecting a mask range of all possible numbers:
\begin{theorem}[Perfectly Secure Masking Schemes]\label{thm:perfect-secure-masking}
If the masking scheme encodes all values in a finite field and selects the mask $\epsilon_i$ uniformly randomly from all field elements, then such a scheme is perfectly secure with 0\% failure rates. The observed value $(x_i + \epsilon_i)$ will uniformly randomly distribute across the field elements regardless of which $x_i$ is selected. 
\end{theorem}

\newcommand{\macroFailureRate}[1]{\ensuremath{\frac{2 #1}{2 \epsilon + #1}}\xspace}

\Cref{thm:perfect-secure-masking} describes the perfect masking scheme in which the adversary \adversarySymbol{} can only observe uniformly random values. However, this scheme requires finite field operations, which are computationally expensive. Instead, we extend this masking mechanism to floating point numbers with bounded failure rates as follows: 
\begin{theorem}[Failure Rates for $\epsilon$-Masking]\label{thm:failure-rates-masking}
    Assuming $x_\delta = | x_1 - x_2 | $ and $\epsilon = k\, x_\delta \ (k \geq 1)$. If \clientSymbol{} selects mask $\epsilon_i \in [-\epsilon, \epsilon]$ instead of using finite field elements, then \adversarySymbol{}, after observing a masked value of $(x_i + \epsilon_i)$, will be able to distinguish the original value between $x_1$ and $x_2$ with a probability of $\frac{2}{2k+1}$.
\end{theorem}

\Cref{thm:failure-rates-masking} describes the probability of an adversary defeating the indistinguishability property of a masking scheme (\Cref{def:masking-failures}). We can apply this failure rate calculation to our solutions for data privacy and model confidentiality with the following conservative estimates:

\begin{corollary}[Estimated Failure Rates for Data Privacy and Model Confidentiality]\label{cor:failure-rates-privacy-confidentiality}
Assuming input data $x = (x_1, x_2, \\ \cdots, x_n)$, $x_\delta = \max(x) - \min(x)$, and $\epsilon = k \, x_\delta \ (k \geq 1)$. Then $\forall x_i \in x$, we estimate the failure rate of applying $\epsilon$-masking is $\frac{2}{2k+1}$. We expect the masking scheme to leak up to $\frac{2n}{2k+1}$ data points from $x$ overall. Similar estimates also apply to model weights $w$ and biases $b$.
\end{corollary}

\def\serverSymbol{\ensuremath{\mathcal{S}}\xspace}

\paragraph{Inference Integrity}
First, we define the offloading scheme with partial verification as follows:
\begin{definition}[$\alpha$-Verified Offload]\label{def:partial-verification-offloading}
    Assuming client \clientSymbol{} wants to offload the computation of function $\mathcal{F}: \mathbb{R}^m \rightarrow \mathbb{R}^n$ to server \serverSymbol{} using the following steps:
    \begin{itemize}
        \item \clientSymbol{} shares input data $x = (x_1, x_2, \cdots, x_m)$ with \serverSymbol{}.
        
        \item \serverSymbol{} computes $\hat{y} = (\hat{y}_1, \hat{y}_2, \cdots, \hat{y}_n)$ and shares $\hat{y}$ with \clientSymbol{}.
        \vspace{-2mm}
        \item \clientSymbol{} selects a verification set $V \subseteq \hat{y}$ such that $|V| = \lceil \alpha \, n \rceil, \alpha \in (0, 1]$. \clientSymbol{} verifies that $\forall \hat{y}_i \in V, \hat{y}_i = y_i$, where $y = \mathcal{F}(x)$.
    \end{itemize}
\end{definition}

With this definition, we can calculate the failure rates of \verisplit{}'s partial verification mechanism under multiple offloading iterations:
\begin{theorem}[Failure Rates for $\alpha$-Verification over $k$ Iterations]\label{thm:failure-rates-integrity}
    Assuming \clientSymbol{} and \serverSymbol{} conducts $k$ rounds of $\alpha$-verified offload. For each offload, \serverSymbol{} reports partially incorrect results, with ratio $\beta$. The probability of $\alpha$-verification fails to detect incorrect results is $\left(\frac{\binom{n-b}{a}}{\binom{n}{a}}\right)^k$, where $a = \lceil \alpha \, n \rceil, b = \lceil \beta \, n \rceil $, $\binom{.}{.}$ is binomial coefficient (combination).
\end{theorem}

\paragraph{Proofs} We provide formal proofs for previous mentioned theorems and corollaries.

\newcommand{\fieldZ}{\ensuremath{\mathbb{Z}_p}\xspace}

\begin{proof}[Proof of \Cref{thm:perfect-secure-masking}]
All numbers (e.g., $x_1$, $x_2$) are embedded in a finite field \fieldZ{}.
Assuming client \clientSymbol{} picks masking noise $\epsilon_i$ uniformly randomly from finite field \fieldZ{}. After applying the mask, \clientSymbol{} shares the value of $(x_i + \epsilon_i) \in \fieldZ{}$ with the adversary \adversarySymbol{}. \adversarySymbol{} has no advantage in determining the original values of $x_i$ since both $(x_1 + \epsilon_i)$ and $(x_2 + \epsilon_i)$ would appear uniformly random across all field elements in \fieldZ{}. 
\end{proof}

\begin{figure}[t]
    \centering
    \begin{tikzpicture}
    \def\valone{60pt}
    \def\valtwo{100pt}
    \def\mask{60pt}

    \def\lineoneY{0pt}
    \def\linetwoY{-30pt}

    \draw[|-|] (\valone - \mask, \lineoneY) -- (\valone + \mask, \lineoneY) ;
    \draw[|-|] (\valtwo - \mask, \linetwoY) -- (\valtwo + \mask, \linetwoY) ;

    \def\tik{(0pt, 3pt) -- (0pt, -3pt)}
    
    \draw[shift={(\valone, \lineoneY)}, color=black] \tik node[below] 
        {$x_1$};
    \draw[shift={(\valone - \mask, \lineoneY)}] \tik node[below] {$x_1 - \epsilon$};
    \draw[shift={(\valone + \mask, \lineoneY)}] \tik node[above=3pt] {$x_1 + \epsilon$};

    \draw[shift={(\valtwo, \linetwoY)}, color=black] \tik node[below] 
        {$x_2$};
    \draw[shift={(\valtwo - \mask, \linetwoY)}] \tik node[below] {$x_2 - \epsilon$};
    \draw[shift={(\valtwo + \mask, \linetwoY)}] \tik node[below] {$x_2 + \epsilon$};
    
    \draw[draw=orange, pattern=north west lines, pattern color=orange] (\valone - \mask, \lineoneY) rectangle (\valtwo - \mask, \lineoneY + 6pt);

    \draw[draw=orange, pattern=north west lines, pattern color=orange] (\valone + \mask, \linetwoY) rectangle (\valtwo + \mask, \linetwoY + 6pt);

    \draw[dashed, color=black] (\valtwo - \mask, \lineoneY) -- (\valtwo - \mask, \linetwoY);
    \draw[dashed, color=black] (\valone + \mask, \lineoneY) -- (\valone + \mask, \linetwoY);
    
\end{tikzpicture}
    \caption{Visualization of masking failure probabilities. If the observed value ($x_i + \epsilon_i$) falls within the orange region, adversary \adversarySymbol{} can distinguish the original value.}
    \label{fig:visualize-failure-rates}
\end{figure}
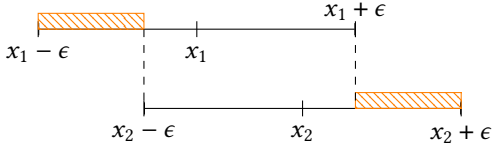

\begin{proof}[Proof of \Cref{thm:failure-rates-masking}]
    Recall $x_{\delta} = | x_1 - x_2 |$ and $\epsilon = k \, x_{\delta} \ (k \geq 1)$. Assuming, without loss of generality, that $x_1 \leq x_2$. \Cref{fig:visualize-failure-rates} provides an example visualization. If the masked value $(x_i + \epsilon_i)$ falls between the regions $[x_1 - \epsilon, x_2 - \epsilon]$ and $[x_1 + \epsilon, x_2 + \epsilon]$ (orange regions in the figure), then \adversarySymbol{} can distinguish whether the original value is $x_1$ or $x_2$. Otherwise, \adversarySymbol{} has no advantage over randomly guessing with a 50\% success rate. Therefore, the failure rate of this $\epsilon$-masking scheme is:

    \begin{align*}
        \text{failure rates} &= \frac{(x_2 - \epsilon) - (x_1 - \epsilon) + (x_2 + \epsilon) - (x_1 + \epsilon)}{(x_2 + \epsilon) - (x_1 - \epsilon)} \\
        &= \frac{x_2 - x_1 + x_2 - x_1}{x_2 - x_1 + 2\epsilon} = \frac{2}{2k+1} \qedhere
    \end{align*}
\end{proof}

\begin{proof}[Proof of \Cref{cor:failure-rates-privacy-confidentiality}]
    We assume there are $n$ elements in data $x = (x_1, x_2, \cdots, x_n)$ (similar proof holds for weights $w$ and biases $b$). For any individual data point $x_i \in x$, the \emph{worst-case} estimated failure rates are bounded by the minimal and maximal values in $x$ (hence, $x_\delta$). Therefore, the upper bound of failure rates for \emph{individual} data points is $\frac{2}{2k+1}$ (\Cref{thm:failure-rates-masking}). Because $\epsilon$-masking scheme is applied to each data point \emph{independently}, the total expected number of failed points in $x$ is $\frac{2n}{2k+1}$ in the worst case. 
\end{proof}

\begin{proof}[Proof of \Cref{thm:failure-rates-integrity}]
    First, consider the failure probability of $\alpha$-verification in a single round of offload. Let $a = \lceil \alpha \, n \rceil = |V|$ and $b = \lceil \beta \, n \rceil$. With $\alpha$-verification, \clientSymbol{} fails to detect misbehaving \serverSymbol{} if and only if no points in $V$ selected by \clientSymbol{} are incorrect. Therefore, by definition of combination, we calculate the failure probability of a single round offload with verification as
    $ \frac{\binom{n-b}{a}}{\binom{n}{a}} $
    : the total number of ways of choosing $a$ points from correct values in $\hat{y}$, divided by the total number of ways of choosing $a$ points from all values in $\hat{y}$. 

    Similarly, for $k$ iterations, \serverSymbol{} must consistently avoid detection by \clientSymbol{}. Therefore, the overall failure rate for $k$ iterations is the product of individual failure rates:
    $ \left( \frac{\binom{n-b}{a}}{\binom{n}{a}} \right)^k $
\end{proof}

\section{Floating Point Mask Precisions}
\label{sec:floating-point}

As explained in \Cref{sec:security-analysis}, \verisplit{} uses floating point numbers to provide security guarantees without converting models into finite fields or using quantization. Unfortunately, this approach introduces new challenges due to precision issues. 


\begin{figure}[t]
    \centering
    \includegraphics[width=0.33\textwidth]{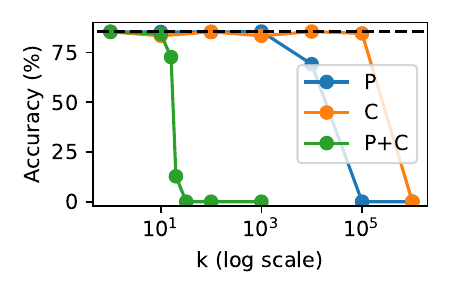}
    \vspace{-5mm}
    \caption{Inference accuracy of ImageNet data after applying different magnitudes of masks to data (for \textbf{\underline{P}}rivacy) and model weights (for \textbf{\underline{C}}onfidentiality). Neither option has noticeable impacts unless $k \geq 10^4$. However, combining both quickly degrades inference accuracy when $k > 10^1$. }
    \label{fig:mask-precision}
\end{figure}

With floating points masks, the key security parameter is $k$: the scale/multiplier of masks relative to input ranges (i.e., $\epsilon = k x_\delta $, \Cref{thm:failure-rates-masking}). As $k$ gets larger, the failure rate ($\frac{2}{2k+1}$) gets smaller and smaller. Ideally, we would make $k$ as large as possible. However, we quickly run into a limitation of floating point operations: precision issues. If the mask value is too large, it is impossible to recover original results from the masked values (e.g., $ y = y' - W \epsilon_i $ in \Cref{fig:workflow-data-privacy}) due to a loss in precision ($y' \approx W\epsilon_i$). Therefore, we must consider practical limits on how large $k$ can be.

\Cref{fig:mask-precision} measures the inference accuracy of the Vision Transformer model (ViT-L16) on the ImageNet validation dataset using different masking parameters. Applying masks for either data privacy or model confidentiality alone does not affect accuracy significantly for smaller $k$ values ($<10^4$). Eventually, masks become too large and imprecise for \verisplit{} to produce accurate inference results. Moreover, combining masking for both privacy and confidentiality significantly limits the maximum values of $k$ before losing precision. We observe a drop of accuracy for $k$ between $10^1$ and $10^2$. This is because of the multiplicative nature of masking (i.e., applying masks for both $W$ and $x$ and computing $Wx$). 

Setting $k = 10$ may seem like a relatively weak security guarantee. However, our security analysis is based on the indistinguishability of two values (\Cref{thm:failure-rates-masking}); hence, the failure rates are a conservative estimate for an attacker to recover original values based on masked observations. To understand the practical implications of these parameters, we design an empirical attack simulation to measure the recovery error rates under various $k$'s. Specifically, we define a game where client \clientSymbol{} chooses values and adversary \adversarySymbol{} tries to recover the originals with minimal errors (in $L_1$ distance):
\begin{definition}[Attack on Floating Point Masks]\label{def:floating-point-mask-attack}
The attack on masked values is conducted as follows:
\begin{itemize}
    \item \clientSymbol{} uniformly randomly selects a value $x_i \in [x_{min}, x_{max}]$ and a mask $\epsilon_i \in [-\epsilon, \epsilon]$, where $\epsilon = k (x_{max} - x_{min})$. 
    \item \adversarySymbol{} receives the value $x_m = (x_i + \epsilon_i)$ and the range $[x_{min}, x_{max}]$. \adversarySymbol{} tries to guess $x'$ with two strategies. The first one is to guess $x'_i \in [x_{min}, x_{max}]$ randomly as a baseline. The second one is to guess $x'_i \in [x_m - \epsilon, x_m + \epsilon]$ (clipped by $x_{min}$, $x_{max}$). 
    \item We measure the average errors as $|x_i - x'_i|$. 
\end{itemize}
\end{definition}

\begin{figure}[t]
    \centering
    \includegraphics[width=0.3\textwidth]{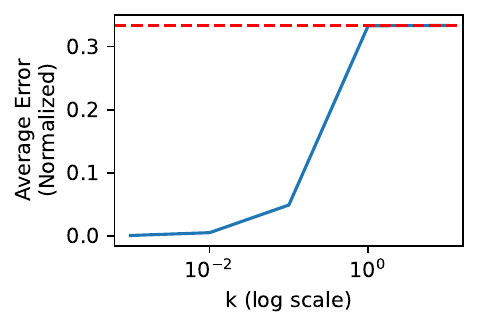}
    \vspace{-2mm}
    \caption{Average recovery errors for attackers to predict original value $x$ based on observed masked value $x'$. The red dashed line indicates randomly guessing values from $[x_{min}, x_{max}]$. The blue line represents guessing values from $[x'-\epsilon, x'+\epsilon]$ where $\epsilon$ depends on the masking scale multiplier $k$.}
    \label{fig:simulate-error-rates}
\end{figure}

\Cref{fig:simulate-error-rates} presents the average errors empirically measured following \Cref{def:floating-point-mask-attack}. We normalize errors by the range of the input ($x_{max} - x_{min}$). The red dashed line represents the baseline attack of uniformly randomly guessed values from $[x_{min}, x_{max}]$. The blue line represents the more sophisticated strategy of $\epsilon$-bounded guesses. When $k$ is small ($\leq 10^{-1}$), \adversarySymbol{} has a strong advantage in recovering the original values with low errors. However, as $k$ gets larger ($>1$), the adversary has no optimal strategy compared to random guesses. These results demonstrate that our security analysis (\Cref{thm:failure-rates-masking}) is very conservative, and $k$ doesn't have to be too large to effectively protect masked data.

\section{Implementation}
We implemented a prototype of \verisplit{} in \codebaseLinesOfCode{} lines of Python code. We implemented \verisplit{}'s machine learning module based on TensorFlow~\cite{website-tensorflow} framework and communication serialization using Google Protocol Buffers~\cite{website-protobuf}. We implemented the Merkle tree construction and verification operation using \texttt{pymerkle} Python library~\cite{website-pymerkle}. 
In \verisplit{} we assume all workers are connected to wall power and hence do not consider battery constraints. 
We also explored alternative embedded devices with TPU accelerators (e.g., Pixels with Tensor chips~\cite{website-pixel-tensor} and Coral USB Edge TPU~\cite{website-coral}). Unfortunately, TPUs (and their TensorFlow Lite SDK~\cite{website-tf-lite}) lack high precision floating point support and require model quantization~\cite{website-coral-overview} and quantization-aware training to provide competitive behavior~\cite{nagel2021quantization-whitepaper}. This violates our practical goals of maintaining high flexibility with floating point operations (\Cref{sec:design-goals}), so we did not pursue this route.

We implemented several performance optimizations, including model weight in-memory caching, multi-threading, and data compression. These optimizations bring mixed returns. For holistic offloading, they greatly reduced inference and verification latency. However, layer-by-layer offloading and memory-constrained devices experience negative results as these options exacerbate resource contention. For example, sending uncompressed data from memory-constrained Raspberry Pi has lower latency due to lower CPU utilization.

\section{Evaluation}
\label{sec:evaluation}

We evaluate the performance overhead of \verisplit{} with various security options compared to local execution. 
We present results from two types of machine learning models --- Vision Transformers~\cite{dosovitskiy2021vit} and a convolution-based VGG16~\cite{simonyan2014vgg}. Their architectural differences lead to different practical choices (layer-by-layer vs. holistic).

\subsection{Setup}
\label{sec:eval-setup}
We use a 4-core Raspberry Pi 4 (RPi4, 64 bit ARM Cortex-A72 cores, running at 1.5Ghz) and 4GB RAM as an example IoT device. To keep a representative setup, we manually restrict RPi's available resources for the \verisplit{} application, considering 1) many IoT devices are less powerful (e.g., using micro-controllers~\cite{teardown-lifx, teardown-teckin}) and 2) similar-equipped devices must share resources across many applications~\cite{news-roomba-qualcomm-chips, boovaraghavan2023mites}. We enforce hard limits on memory capacity and processors by modifying the Linux boot configuration. We allocate 4GB for memory swap space. The RPi connects to a local LAN either over Gigabit Ethernet or the 5GHz WiFi interfaces, depending on the evaluation.

For the local worker, we choose a gaming desktop (16 cores, 48GB RAM) with a GPU (RTX 3090, 24GB VRAM). The computer is connected to the LAN via Gigabit Ethernet. During our evaluation, the worker does not execute any resource-intensive workload and remains idle. To emulate multiple workers for model confidentiality, we create multiple \verisplit{} runtime on the same worker machine. 

We selected the ImageNet~\cite{ILSVRC15imagenet} dataset for offloading workload and measured the average inference latency over \iters{} inferences. The offloading device conducts each inference in single batch to better capture real-time inference experience. We incorporate two representative machine learning models --- a classic, convolution-based one (VGG16~\cite{simonyan2014vgg}) and a more recent, attention-based family of Vision Transformers~\cite{dosovitskiy2021vit} (different sizes and configurations).

\subsection{Vision Transformers}
\label{sec:eval-vit}

In this section, we present the evaluation results for offloading Vision Transformer (ViT) models with \verisplit{}. We employ a \emph{layer-by-layer} offloading approach to demonstrate \verisplit{}'s full capability of protecting data privacy, model confidentiality, and inference integrity. 

For ViT models, we offload computationally-expensive dense layers in MLP to workers while keeping other layers (e.g., non-linear layers, attentions) on the device. This reduces computation demands on the device without incurring too much communication overhead. 

\begin{figure*}[t]
\centering
\begin{subfigure}{0.24\textwidth}
\includegraphics[width=\textwidth]{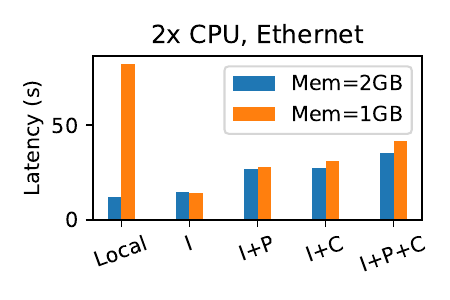}
\vspace{-8mm}
\caption{Impact of Memory Size}\label{fig:offloading_overhead:memory}
\end{subfigure}
\hfill
\begin{subfigure}{0.24\textwidth}
\includegraphics[width=\textwidth]{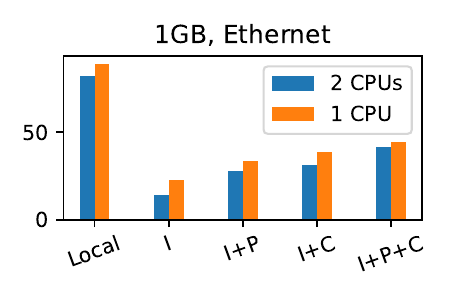}
\vspace{-8mm}
\caption{Impact of CPU Cores}\label{fig:offloading_overhead:cpus}
\end{subfigure}
\hfill
\begin{subfigure}{0.24\textwidth}
\includegraphics[width=\textwidth]{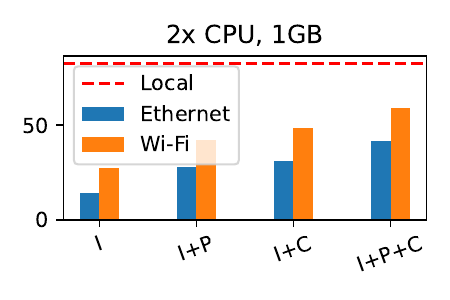}
\vspace{-8mm}
\caption{Impact of the Network}\label{fig:offloading_overhead:ethernet-wifi}
\end{subfigure}
\hfill 
\begin{subfigure}{0.24\textwidth}
    \includegraphics[width=\textwidth]{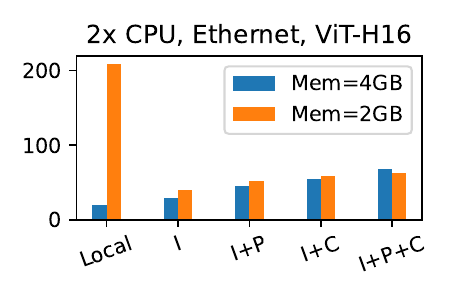}
    \vspace{-8mm}
    \caption{Memory Impact (ViT-H16)}\label{fig:offloading_overhead:vit_huge}
\end{subfigure}
\caption{Average inference latency of Vision Transformer models (ViT-L16 unless noted otherwise) with guarantees: Integrity (I), Data Privacy (P), and Model Confidentiality (C), under different configurations for the offloading device.
(\subref{fig:offloading_overhead:memory}) Latency comparison of memory sizes.
(\subref{fig:offloading_overhead:cpus}) Latency comparison of numbers of CPU cores.
(\subref{fig:offloading_overhead:ethernet-wifi}) Latency comparison of different network conditions.
(\subref{fig:offloading_overhead:vit_huge}) Latency comparison of a larger model (ViT-H16) with different memory sizes.
}
\label{fig:offloading_overhead}
\end{figure*}

\paragraph{Inference Latency}
\Cref{fig:offloading_overhead} presents the average latency of model inferences with the ViT Large model (ViT-L16, with a batch size of 16) under various settings. As a baseline, we select local execution on the IoT device rather than comparison with first-party cloud offloading, which would require vendors to set up an additional cloud backend.

\newcommand{\vitOneGBbaseline}{82\xspace}
\newcommand{\vitOneGBOffloadReduction}{49\%--83\%\xspace}

First, offloading with \verisplit{} outperforms local execution for devices with limited computation. \Cref{fig:offloading_overhead:memory} shows that, for a device with 1GB memory, running ViT-L32 locally takes \vitOneGBbaseline{} seconds, while \verisplit{} offloading reduces latency by \vitOneGBOffloadReduction{}, depending on the set of options enabled. One might argue that the IoT device could increase memory to 2GB, and the local execution would improve significantly. However, as machine learning models get more complex, larger models would quickly exceed device hardware capability. For example, even 2GB memory is insufficient for a larger vision transformer (ViT-H16, \Cref{fig:offloading_overhead:vit_huge}), which provides better accuracy. Looking forward, we believe \verisplit{} provides a practical solution to enable the execution of complex large models on resource-limited IoT devices with strong security and privacy guarantees.

In addition to memory size, we observe a slight impact from the number of CPU cores. \Cref{fig:offloading_overhead:cpus} presents the average inference latency on a device with 1GB memory and connected over Ethernet. Increasing CPU cores from 1x to 2x reduces latency of \verisplit{} offloading by up to 37\% using the same set of options. In comparison, local execution only experiences $10\%$ latency reduction due to it being memory constrained (i.e., not bounded by CPU).

As expected, different network conditions significantly impact the performance of \verisplit{} offloading. \Cref{fig:offloading_overhead:ethernet-wifi} compares the average inference latency between WiFi 5GHz and Ethernet connections for devices with 2 CPU cores and 1GB memory. The horizontal red dashed line indicates the local execution baseline. Switching the connection from Ethernet to WiFi raises the inference latency by up to 1.89x. However, it is important to note that offloading over WiFi with all options enabled still outperforms local inferences. 

\begin{table}[t]
    \centering
    \begin{tabular}{P{10mm}|P{25mm}|P{15mm}|P{15mm}}
     \textbf{CPU (cores)} & \textbf{Pre-Process (seconds/mask)} & \textbf{Inference (seconds)} & \textbf{Verification (seconds)} \\
     \hline
     1 & \FPeval{\result}{round(724/\iters{},2)}$\result$ & \FPeval{\result}{round(1296/\iters{},2)}$\result$ & 17 \\
     \hline 
     {2} & \FPeval{\result}{round(660/\iters{},2)}$\result$ & \FPeval{\result}{round(1402/\iters{},2)}$\result$ & 12.36 \\
\end{tabular}
    \caption{Pre-process and verification time for \verisplit{} with data privacy and integrity options, measured on a device with 1GB memory. Both overheads can be asynchronous: generate masks before inference starts and verify anytime after inference finishes.}
    \label{tab:pre-processing-time}
\end{table}

\paragraph{Additional Overhead}
In addition to inference delays, enabling different options for \verisplit{} incurs additional overhead, such as pre-processing for data privacy and verification to ensure integrity. \Cref{tab:pre-processing-time} presents these overheads compared to inference times. Here the device verifies \emph{all} results without the need for partial verification. As expected, more CPU cores help reduce latency, although they both take considerable time. Unlike time-sensitive inferences, the device can pre-process multiple masks during idle times for future use and aggregate verification for many prior inferences altogether. Therefore, these delays should not impact real-world deployment performance too much. Even with both overheads combined, \verisplit{} with data privacy and integrity verification still outperforms local inferences (\Cref{fig:offloading_overhead:cpus}).

\subsection{Convolutional Neural Networks}
\label{sec:eval-vgg}

In addition to Vision Transformers, we evaluate \verisplit{} with a more traditional, convolution-based VGG16 model. Unlike transformers, VGG16 utilizes convolution layers extensively, which generate large amounts of intermediate values (layer inputs and outputs) as measured by prior research~\cite{kang2017neurosurgeon, tramer2018slalom}. Therefore, offloading VGG16 layer-by-layer incurs high communication overhead, limiting the practicality of \verisplit{} and the performance benefits. Instead, we employ a \emph{holistic} offloading approach to ensure inference integrity with \verisplit{} with minimal overhead. This implies \verisplit{} can guarantee inference integrity in this case. 

\begin{figure*}[t]
    \centering
    \includegraphics[width=0.85\textwidth]{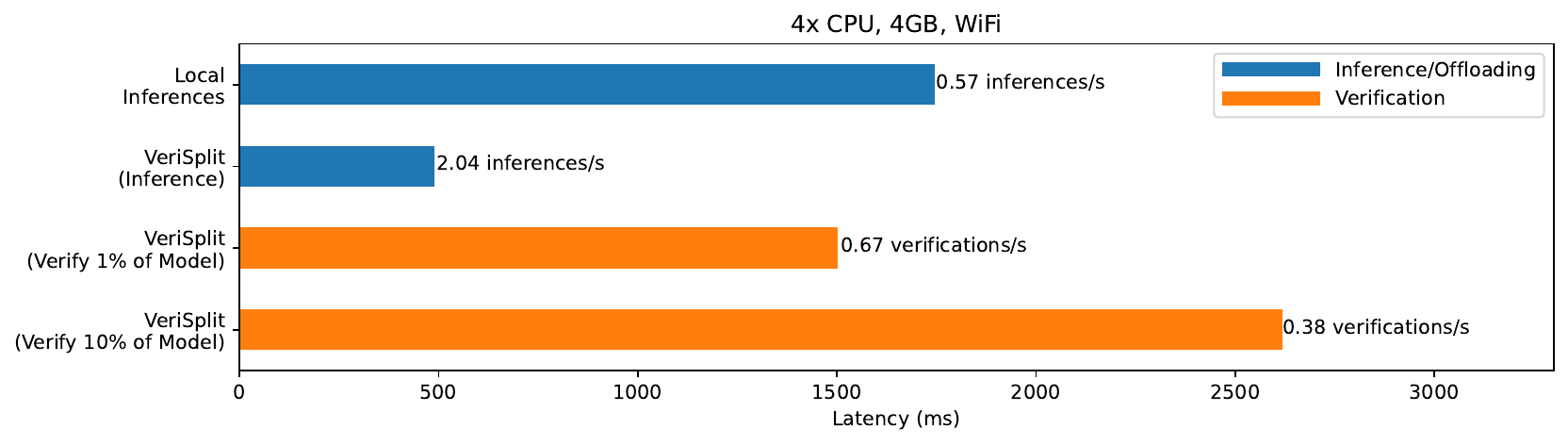}
    \caption{Latency for offloading and verifying VGG16 model with \verisplit{}. The local execution baseline provides 0.57 inference per second, while integrity-enabled \verisplit{} can provide 2.04 inference/s. Notably, for \verisplit{}, verification can be performed asynchronously when the device is idle. Verification overhead includes the workers assembling the proofs, transmitting them over the network (WiFi in this setting), and the device recomputing the results and validating the proofs.}
    \label{fig:vgg-latency-breakdown}
\end{figure*}

\newcommand{\vggBaseline}{1746\xspace}
\newcommand{\vggInference}{\FPeval{\result}{round(134.0557769+45.64143426+230.7370518+80.20717131,0)}\result\xspace}

\Cref{fig:vgg-latency-breakdown} presents the average latency for inferences with VGG16 models with integrity verification. We evaluate these tasks on a device with 4x CPU cores and 4GB memory. As a baseline, it takes \vggBaseline{} ms to perform inferences locally. The device can use \verisplit{} to offload inferences to a worker with a GPU. This helps reduce inference latency to \vggInference{} ms. The most significant benefit of \verisplit{} offloading, in this case, is reduced CPU loads on the device (from 2.61 for local inferences to 0.63 for \verisplit{} with verification, not shown in the figure). If the device selects 1\% of the intermediate values to verify, the verification overhead averages around 1503 ms, but this step can happen asynchronously. If the device wants to verify more values, it must spend more time collecting intermediate results and validating proofs.

\section{Limitations and Discussion}
\label{sec:discussion}

\paragraph{Applying \verisplit{} to Cloud Offloading}
We designed \verisplit{} as a solution for secure and private offloading across local IoT devices. Therefore, we made a few design decisions based on network conditions and trade-offs between communication and computation overhead. We believe that the techniques we propose in \verisplit{} should also apply to certain cloud offloading applications depending on the network condition. For example, data-center networking may provide higher bandwidth and lower latency, making it ideal for \verisplit{} solutions. However, offloading over a WAN might introduce significantly higher latency and potentially metered bandwidth, and is thus less applicable for \verisplit{}. 

\paragraph{Extending \verisplit{} to Additional Models}
Incorporating new models into \verisplit{} requires a few steps. First, analyzing whether layer-by-layer or holistic offload is more suitable is important. If the new model involves layers with large intermediate values (e.g., convolution layers), offloading those layers might not be worthwhile due to the communication overhead. Therefore, one task is determining where each layer should be executed (locally or remotely) and how many layers should be offloaded altogether. Second, for new layer types, it is necessary to implement solutions for verifying partial results (instead of running the entire layer) to save on computation costs.

\section{Conclusion}
In this paper, we propose an efficient solution to securely and privately offload deep models across IoT devices, improving the utilization of local idle resources while reducing operational costs. We introduce \verisplit{}, a novel offloading framework design, to address several practical challenges in protecting data privacy, model confidentiality, and inference integrity for resource-constrained IoT devices. We propose masking mechanisms to protect privacy and confidentiality by adding noises to their real values. We introduce an inference commitment mechanism based on Merkle trees to separate verification steps from the real-time inference process. In addition, we address challenges in floating-point computation errors and precision issues compared to prior approaches relying on finite fields. Our prototype evaluation results demonstrate \verisplit{}'s practicality and low overhead.

\clearpage
\bibliographystyle{plain}
\bibliography{bibliography}

\end{document}